\documentclass[pre,twocolumn]{revtex4}
\usepackage[dvips]{graphicx}
\usepackage{amsmath}

\newcommand{\avg}[1]{\ensuremath{\langle #1 \rangle}}

\begin{document}

\title{Vicious walkers in a potential}
\date{\today}

\author{Alan J.\ Bray}
\author{Karen Winkler}
\affiliation{Department of Physics and Astronomy, University of 
Manchester, Manchester M13 9PL, U.K.}
 
\begin{abstract}
We consider $N$ vicious walkers  moving in one dimension in a one-body
potential $v(x)$. Using the  backward Fokker-Planck equation we derive
exact  results for  the asymptotic  form of  the  survival probability
$Q({\bf  x},t)$  of  vicious   walkers initially located at 
$(x_1,\ldots,x_N)={\bf x}$,  when  $v(x)$  is  an  arbitrary
attractive  potential. Explicit  results are  given for  a square-well
potential  with absorbing  or  reflecting boundary  conditions at  the
walls, and  for a harmonic  potential with an absorbing  or reflecting
boundary at the  origin and the walkers starting  on the positive half
line. By mapping the problem  of $N$ vicious walkers in zero potential
onto  the  harmonic potential  problem,  we  rederive  the results  by
Fisher  [J.\ Stat.\ Phys.\ {\bf 34}, 667 (1984)] and   Krattenthaler  
et  al.~[J.\ Phys.\ A {\bf 33}, 8835 (2000)] respectively  for  vicious  
walkers  on  an infinite  line  and  on  a semi-infinite line with an 
absorbing wall at the origin.  This mapping also gives  a new result 
for  vicious walkers on  a semi-infinite line
with  a   reflecting  boundary   at  the  origin:   $Q({\bf  x},t)\sim
t^{-\frac{N(N-1)}{2}}$.
\end{abstract}

\maketitle


\section{Introduction}
Consider $N$ symmetric random walkers which annihilate on meeting each
other  but otherwise  do not  interact. This  concept  of short-ranged
interacting  random walkers  was introduced  by  M. E.  Fisher as  the
vicious walkers model \cite{fisher}.

One of the  main properties of interest in this  model is the survival
probability $Q({\bf  x},t)$ that none of the $N$ vicious walkers with 
initial position coordinates  ($x_{1},x_{2},\dots,x_{N})={\bf x}$ has  
met another up to  time $t$, i.e.\ none of them has  been annihilated 
up to time $t$.

Fisher and Huse~\cite{fisher,husefisher} determined the survival
probability for  $N$ vicious walkers  moving on an infinite  line. For
large times, $Q({\bf x},t)$ decays as a power:
\begin{equation}
Q({\bf x},t)\sim t^{-\frac{N(N-1)}{4}}\ .
\end{equation} 
Interesting results also arise  when further conditions are imposed on
the  movement  of the  vicious  walkers by  the  use  of absorbing  or
reflecting walls, where all walkers  are initially located on the same
side of the  boundary (the case where there are  walkers on both sides
decouples     into    two    independent    problems).    While
Fisher~\cite{fisher} found the survival probability problem of vicious
walkers  with an  absorbing boundary  at  the origin  only for  $N=2$,
Krattenthaler  et al.\  ~\cite{krattenthaler} were  able  to determine
exact  asymptotic   forms  for  $N$  vicious   walkers  starting  from
equi-spaced lattice points:
\begin{equation}
Q({\bf x},t)\sim t^{-\frac{N^{2}}{2}}.
\end{equation} 
By  evaluating  the scaling  limit  Katori and  Tanemura~\cite{katori}
showed that  this asymptotic behaviour holds for  arbitrary initial
conditions on a continuous line.

In  this paper  we introduce  the interesting  problem of  $N$ vicious
walkers moving  in an attractive one-body potential  $v(x)$, i.e.\ the
full   potential    function   has   the    separable   form   $V({\bf
x})=\sum_{i=1}^{N}v(x_{i})$.    Treating  both   time  and   space  as
continuous,  we investigate  the survival  probability of  $N$ vicious
walkers with  equal diffusion constants  $D$.  The equation  of motion
for walker $i$ is taken to be
\begin{equation}
\dot{x}_i = - \frac{\partial V}{\partial x_i} + \eta_i(t),
\label{langevin}
\end{equation}
where the  Langevin noise $\eta_i(t)$  is a Gaussian white  noise with
mean zero and correlator
\begin{equation}
\avg{\eta_i(t) \eta_j(t^\prime)} = 2 D \delta_{ij} \delta(t-t^\prime).
\label{noise}
\end{equation}
For a square-well  potential of width $L$ we  consider three different
combinations of absorbing and reflecting walls and find an exponential
decay  for  the  survival  probability  of the  general  form  $Q({\bf
x},t)\sim  e^{-\theta_{N}t}$.  For two  reflecting walls  the exponent
$\theta_{N}$ is determined to be
\begin{equation}
\theta_{N}^{RR}=D\frac{\pi^{2}}{L^{2}}\frac{N(N-1)(2N-1)}{6}.
\end{equation}
In the case of one reflecting and one absorbing wall we obtain
\begin{equation}
\theta_{N}^{RA}=D\frac{\pi^{2}}{L^{2}}\frac{N(2N+1)(2N-1)}{12},
\end{equation}
while for two absorbing walls the exponent of the asymptotic decay is:
\begin{equation}
\theta_{N}^{AA}=D\frac{\pi^{2}}{L^{2}}\frac{N(N+1)(2N+1)}{6}.
\end{equation}   
An interesting  potential which  turns out to  a powerful tool  is the
problem  of $N$  vicious  walkers in  the  harmonic potential  $V({\bf
x})=\frac{a}{2}{\bf x}^{2}$. The  asymptotic behaviour for large times
is determined to be an  exponential decay independent of the diffusion
constant:
\begin{equation}
\theta_{N}=\frac{N(N-1)}{2}a.
\end{equation} 
This  result  also provides  a  mechanism  to  determine the  survival
probability of  $N$ vicious  walkers on an  infinite line in  a simple
way. By  mapping the zero-potential problem to  the harmonic potential
problem, we  derive Fisher's result~\cite{fisher} and  also the result
by Krattenthaler et al~\cite{krattenthaler}  with an absorbing wall at
the origin. Furthermore we are able to obtain, to our knowledge, a new
result  for the  survival  probability  of $N$  vicious  walkers on  a
semi-infinite  line with a  reflecting boundary  at the  origin, which
decays as:
\begin{equation}
Q({\bf x},t)\sim t^{-\frac{N(N-1)}{2}}.
\end{equation} 

The outline of  the paper is as follows. In section  II we present the
method for a  general one-body potential $v(x)$, while  in section III
we give  explicit results for square-well  and harmonic potentials.
In section  IV we  revisit the case  of zero potential,  obtaining the
known  results,  and a  new  result for  a  system  with a  reflecting
boundary, through a transformation  to the harmonic problem. Section V
is a short conclusion.

\section{The method}
The dynamics of a random walker, with position coordinate $x_i$, moving 
in a potential $V({\bf x})$ is  described, in  continuous space  and 
time,  by the Langevin equation (\ref{langevin}) with noise correlator 
(\ref{noise}). 

The probability $Q({\bf x},t)$ that all $N$ vicious walkers, $i=1,\ldots,N$,  
have survived up to time $t$, given  that they {\em started} at 
\{${x}_i$\}, satisfies the corresponding backward Fokker-Planck equation: 
\begin{equation}
\label{bfp}
\frac{\partial Q({\bf x},t)}{\partial t} 
= D \sum_{i=1}^{N}\frac{\partial^2}{\partial x_{i}^2}Q({\bf x},t)
-\sum_{i=1}^{N}\frac{\partial V({\bf x})}{\partial x_{i}}
\frac{\partial Q({\bf x},t)}{\partial x_{i}}.
\end{equation}
For convenience we start by defining the survival probability $q(x_{i},t)$ 
of just one random walker moving in a potential restricted by the imposed 
boundary conditions. This survival probability $q(x_{i},t)$ satisfies 
the backward Fokker-Planck equation 
\begin{equation}
\frac{\partial q(x_{i},t)}{\partial t} 
= D \frac{\partial^2}{\partial x_{i}^2}q(x_{i},t)
-\frac{dv(x_i)}{dx_{i}}\frac{\partial q(x_{i},t)}{\partial x_{i}},
\label{BFPE}
\end{equation}
where we  have used the  relation $V({\bf x})  = \sum_i v(x_i)$  for a
one-body potential. For any such potential, the backward Fokker-Planck
equation  (\ref{BFPE}) is separable  in time  and space. 
Let us call  these separable solutions, i.e.\  the solutions of 
Eq.\  (\ref{BFPE})  satisfying  the  relevant   boundary  
conditions, single-walker basis functions. They have the form 
$q_j(x_i,t) = u_j(x_i)\exp(-\lambda_j t)$, where $\lambda_j$ is the decay 
rate associated with basis function $j$, and these rates are ordered such 
that $\lambda_1 < \lambda_2 < \lambda_3 \ldots$. 

For $N$ non-interacting walkers moving in the same potential, the 
$N$-walker basis functions for the survival probability take the form 
of products of $N$ single-walker functions, each with a different 
space variable $x_{i}$. Since, however, we are investigating  
vicious walkers  the mutual 
annihilation  property  must be respected. Since two walkers  die when  
arriving   at  the   same $x$-coordinate, the boundary  condition 
$Q(x_1,\ldots,x_n,t) = 0$ when $x_i=x_j$  for any  $i  \ne j$  must  be 
respected.  This property is ensured   by   constructing   $Q({\bf   x},t)$  
using  antisymmetric combinations of products of $N$ single-walker 
functions,  analogous  to  the antisymmetric   construction   of   the   
wavefunction  of  fermions \cite{fisher}. The $N$-walker basis functions 
of the vicious walker problem with $N$ walkers have, therefore, the form
\begin{equation}
\label{antisym}
Q^{i_1,\ldots,i_N}({\bf x},t)= \det A^{i_1,\ldots,i_N},
\end{equation}
where the elements of the $N \times N$ matrix $A$ are given by
\begin{equation}
A^{i_1,\ldots,i_N}_{nm} = q_{i_n}(x_{m},t).
\label{elements}
\end{equation}
The full solution $Q({\bf x},t)$ is a linear superposition of these basis 
functions with coefficients determined by the initial condition. 

To solve the  problem of  $N$ vicious  walkers in  an arbitrary 
potential, therefore, we need to  find the single-walker basis functions 
$q_j(x_i,t)$ appropriate to the imposed boundary conditions. For an 
absorbing boundary at $x=a$ the functions $q_j(x_{i},t)$ must satisfy
\begin{equation}
q_j(x_{i}=a,t)=0.
\end{equation}
For a reflecting boundary at $x=b$ the boundary condition for the 
backward Fokker-Planck equation is in general
\begin{equation}
{\bf\nabla}Q({\bf x},t)\cdot{\bf\hat n}=0,
\end{equation}
where ${\bf\hat n}$ is normal to the reflecting boundary~\cite{gardiner}. 
In the one-dimensional case, this expression implies
\begin{equation}
\frac{dq_j}{dx_i}\Big|_{x_i=b}=0. 
\end{equation}
Clearly these boundary conditions are also satisfied by the functions 
$Q^{i_1,\ldots,i_N}({\bf x},t)$, 
since the latter is just an antisymmetrised product of single-walker 
basis functions.  

Consider now the late-time limit, $t \to \infty$. Each antisymmetrised 
product in the expression for $Q({\bf x},t)$ contains $N$ different 
relaxation factors $\exp(-\lambda_j t)$. The slowest-decaying term in the 
sum, therefore, is the term in which the relaxation rates are 
$\lambda_1,\lambda_2,\ldots,\lambda_N$. It follows that, asymptotically, 
\begin{equation}
Q({\bf x},t) \propto \det B^{1,2,\ldots,N} \exp(-\theta_N t),
\label{general}
\end{equation}
where $B^{1,2,\ldots,N}$ is just the $N\times N$ matrix with elements 
$B_{nm} = u_n(x_m)$ ($n,m = 1,\ldots,N$), i.e.\ it is constructed using 
the $N$ slowest-decaying single-walker basis functions, and the total 
decay rate is 
\begin{equation}
\theta_N = \sum_{j=1}^N \lambda_j\ .
\label{theta}
\end{equation} 
The following sections provide some applications of this general result.

\section{Results for vicious walkers in a potential}
In this section we discuss two examples of N vicious walkers in a potential 
and determine the decay of the survival probability $Q({\bf x},t)$.
\subsection{The square-well potential}
Consider  a square-well  potential  which has  two  walls of  infinite
potential, one at the  origin and  the other  at  $x=L$, and vanishes 
between the walls.  A vicious walker  restricted to  move between  the walls
satisfies the backward Fokker-Planck equation:
\begin{equation}
\frac{\partial        q(x_{i},t)}{\partial        t}        =        D
\frac{\partial^2q(x_{i},t)}{\partial x_{i}^2}
\end{equation} 
This   equation  can   be   solved  in   general   by  separation   of
variables, which amounts in this case to writing the solution as a spatial 
Fourier series. Different  solutions  result  from  the  various  sets  of
boundary conditions imposed by the property of the walls.
\subsubsection{Two reflecting walls}
For two reflecting walls the spatial derivative of $q(x_{i},t)$ must be 
zero at $x=0$ and $x=L$. In this case, therefore $q(x_{i},t)$ is given by 
Fourier cosine series with basis functions 
\begin{equation}
q_n(x_{i},t) = \exp\left(-\frac{n^{2}\pi^{2}D t}{L^{2}}\right)
\cos\left(\frac{\pi}{L}n x_{i}\right),\ \ \ n=0,1,\ldots
\end{equation}
The survival probability is constructed as a superposition of antisymmetrised
products of these basis functions:
\begin{eqnarray}
Q({\bf x},t) & = & \sum_{i_{1}}\dots\sum_{i_{N}}
C^{i_1,\ldots,i_N} \det A^{i_1,\ldots,i_N} \nonumber \\
&=& \sum_{i_{1}}\dots\sum_{i_{N}}
C^{i_1,\ldots,i_N} \exp\left(-\frac{\pi^{2}Dt}{L^{2}}\sum_{n=1}^{N} 
i_{n}^{2}\right) \nonumber \\
&& \times \det B^{i_1,\ldots,i_N}, 
\end{eqnarray}
where 
\begin{equation}
B^{i_1,\ldots,i_N}_{nm} = \cos\left(\frac{\pi}{L}i_{n}x_{m}\right).
\end{equation}
To  evaluate  the long-time behaviour we keep  only the $N$ longest-lived 
modes, given by the $N$ smallest values, $i=0,1,\ldots,N-1$ of $i_{n}$. 
Using $\sum_{i=0}^{N-1}i^{2}=N(N-1)(2N-1)/6$ we obtain, for the asymptotic 
time-dependence, 
\begin{equation}
Q({\bf                                                        x},t)\sim
\exp\left(-\frac{\pi^{2}Dt}{L^{2}}\,\frac{N(N-1)(2N-1)}{6}\right).
\label{QRR}
\end{equation} 

\subsubsection{One reflecting and one absorbing wall}
For an  absorbing wall at  the origin and  a reflecting wall at $x=L$ the 
boundary conditions are satisfied by a Fourier sine series with 
basis functions
\begin{eqnarray}
q_n(x_{i},t)& = & \exp\left(-\frac{(2n+1)^{2}\pi^{2}D t}{4L^{2}}\right)
\nonumber \\
&& \times\sin\left(\frac{\pi}{2L}(2n+1)x_{i}\right),\ \ n=0,1,\ldots
\end{eqnarray}
Analogous  to the preceding  case  the survival  probability  for all $N$
vicious walkers is constructed and the asymptotic survival probability
for large time  is  evaluated using $\sum_{i=0}^{N-1}(2i+1)^2
= N(2N+1)(2N-1)/3$ to give the asymptotic decay
\begin{equation}
Q({\bf                                                        x},t)\sim
\exp\left(-\frac{\pi^{2}Dt}{L^{2}}\,\frac{N(2N+1)(2N-1)}{12}\right).
\label{QAR}
\end{equation} 
\subsubsection{Two absorbing walls}
In the case of two absorbing walls the basis functions have to vanish at
both at $x=0$  and  at $x=L$. A Fourier sine series is therefore 
appropriate, with basis functions
\begin{equation}
q_n(x_{i},t) = \exp\left(-\frac{n^{2}\pi^{2}Dt}{L^{2}}\right)
\sin\left(\frac{\pi}{L}n x_{i}\right), \ \ n=1,2,\ldots
\end{equation}
This is very similar to the result for two reflecting boundaries, except 
that the spatial functions are sines so the sum begins with $n=1$. 
The large-time behaviour $Q({\bf x},t)$ is given by
\begin{equation}
Q({\bf x},t)\sim \exp\left(-\frac{\pi^{2}Dt}{L^{2}}\,
\frac{N(N+1)(2N+1)}{6}\right).
\label{QAA}
\end{equation}
Before proceeding to the harmonic potential, we note that the inequalities
$2N(N-1)(2N-1)<N(2N+1)(2N-1)<2N(N+1)(2N+1)$, for all $N \ge 1$, imply that 
for a well of given size the decay is fastest with two absorbing boundaries 
and slowest with two reflecting boundaries, as is intuitively clear. 

\subsection{The harmonic potential}
A harmonic potential $V({\bf x})=\frac{a}{2}{\bf x}^{2}$ is considered
for which  the backward  Fokker-Planck equation for the one-walker basis 
function reads
\begin{equation}
\label{hpbfp}
\frac{\partial q(x_{i},t)}{\partial  t} = D \frac{\partial^2}{\partial
x_{i}^2}q(x_{i},t)
   -a\;x_{i}\frac{\partial  q(  x_{i},t)}{\partial
x_{i}}.
\end{equation} 
This equation can be transformed into an imaginary-time Schr\"odinger
equation by the substitution $q(x_i,t)= \exp(ax_i^2/4D)\psi(x_i,t)$  
to give
\begin{equation}
\frac{\partial \psi(x_{i},t)}{\partial  t} = D \frac{\partial^2}{\partial
x_{i}^2}\psi(x_{i},t) +\left(\frac{a}{2}-\frac{a^{2}x_{i}^{2}}{4
D}\right) \psi( x_{i},t).
\end{equation} 
This equation has solutions of the form $\psi(x_{i},t)= e^{-\lambda t}
u(x_{i})$, where $u(x_{i})$ satisfies the ordinary differential equation
\begin{equation}
\left(D \frac{d^2}{dx_{i}^2}+\left(\frac{a}{2}-\frac{a^{2}x_{i}^{2}}
{4D}\right)\right)u(x_{i})=-\lambda u(x_{i}).
\end{equation}
This equation is equivalent to the time-independent Schr\"odinger 
equation for the harmonic oscillator. The  eigenvalues and  
eigenfunctions of this  eigenvalue problem  are well known: see  for 
example a similar problem  in ref.~\cite{reichl}. The eigenfunctions 
have the form 
\begin{equation}
u_n(x_{i})=H_{n}\left(x_i\sqrt{\frac{a}{2D}}\right)\,
\exp(-\frac{a}{4 D}x_i^2),
\end{equation}
where the functions $H_n(x)$ are the Hermite polynomials defined by
\begin{equation}
H_{n}(y)=(-1)^{n}e^{y^{2}}\frac{d^{n}}{dy^{n}}e^{-y^{2}}.
\end{equation}
The corresponding eigenvalues are $\lambda_n = na$, where $n=0,1,2,\dots$.
The original basis functions $q(x_i,t)$ are given by $q_n(x_i,t) = 
H_{n}\left(x_i\sqrt{\frac{a}{2D}}\right)\exp(-\lambda_n t)$.  

Applying  the  antisymmetrisation  process  to  determine  the  survival
probability  of  $N$  vicious  walkers in a harmonic potential we  obtain  
the asymptotic time dependence:
\begin{equation}
Q({\bf x},t)\sim \exp\left(-at\sum_{i=0}^{N-1} i \right)
\end{equation} 
giving
\begin{equation}
Q({\bf x},t)\sim \exp\left(-at\,\frac{N(N-1)}{2}\right).
\label{harmonic}
\end{equation} 

This approach is readily extended to the case where there is a reflecting 
or absorbing boundary at $x=0$ and all the walkers start on the same side 
of the boundary (if there are walkers on both sides, the problem decouples 
into two independent problems). For a reflecting boundary, the boundary 
condition $u'(0)=0$ selects only the even-numbered Hermite polynomials, 
$n=0,2,4,\ldots$, and 
\begin{eqnarray}
Q({\bf x},t) & \sim & \exp\left(-at\sum_{i=0}^{N-1} 2i \right) \nonumber \\
& = & \exp[-at\,N(N-1)]\ ({\rm reflecting\ wall}).
\label{reflecting}
\end{eqnarray}
For an absorbing boundary, the boundary condition $u(0)=0$ selects the 
odd-numbered Hermite polynomials to give
\begin{eqnarray}
Q({\bf x},t) & \sim & \exp\left(-at\sum_{i=1}^{N} (2i-1) \right) \nonumber \\
& = & \exp[-at\,N^2]\ ({\rm absorbing\ wall}).
\label{absorbing}
\end{eqnarray}

In the following section we show how these results can be used to compute 
the survival probability of $N$ vicious walkers in {\em zero} potential, 
with and without an absorbing or reflecting wall, by mapping the problem 
back to the oscillator problem.  

\section{Vicious walkers on a line}
Here  the case  of $N$  vicious walkers  restricted by  no  potential is
investigated.  This  problem can  be solved in  a quite simple  way by
mapping it to the problem of $N$ vicious walkers in a harmonic potential
and using the previous results. Again,  we consider the  Langevin equation
(\ref{langevin}), but with $V({\bf  x})=0$, and  let all $N$ vicious walkers
start to  move at time  $t=t_{0}$. We introduce the  following mapping
from ${\bf x},t$ to the new coordinates ${\bf X},T$ by~\cite{map}:
\begin{equation}
{\bf X}=\frac{{\bf x}}{\sqrt{2Dt}},\qquad t=t_{0}\,e^{T}.
\label{cov}
\end{equation}
Then the Langevin equation (\ref{langevin}) transforms to
\begin{equation}
\frac{dX_i(T)}{dT} = -\frac{1}{2}X_i(T)+\xi_i(T)
\end{equation}
where $\xi_i(T)=\sqrt{t_{0}/2D}\,e^{T/2}\,\eta_i(t_0\,e^T)$ is a
Gaussian white noise with mean zero and correlator
\begin{equation}
\avg{\xi_i(T) \xi_j(T^\prime)} = \delta_{ij} \delta(T-T^\prime).
\end{equation}
The   corresponding  backward  Fokker-Planck   equation  in   the  new
coordinates is
\begin{equation}
\frac{\partial    Q({\bf     X},T)}{\partial    T}    =    \frac{1}{2}
\sum_{i=1}^{N}\frac{\partial^2}{\partial     X_{i}^2}Q({\bf    X},T)
-\frac{1}{2}\sum_{i=1}^{N}X_{i}\frac{\partial  Q({\bf  X},T)}{\partial
X_{i}}
\end{equation}
where the space coordinates are now the starting points of the vicious
walkers, given by:
\begin{equation}
X_{i}(T=0)=\frac{x_i(t_{0})}{\sqrt{2Dt_{0}}}\ .
\end{equation}
In the  new coordinates this  problem looks identical to  the harmonic
potential problem  with $a=1/2$ and  $D=1/2$.  Hence the  asymptotic (in 
time) solution for the  survival probability of $N$ vicious walkers is,  
according to  our previous results,
\begin{eqnarray}
Q({\bf X},T) \sim \exp\left(-\frac{T}{2}\sum_{i=0}^{N-1} i \right)\det B^{H}
\end{eqnarray}
where  $(B^{H})_{nm} = H_{n-1}(X_{m}/\sqrt{2})$ and $n,m=1,\ldots,N$.  
Mapping back  to the original  coordinates $({\bf  x},t)$ leads  to the  
survival probability
\begin{equation}
Q({\bf x},t)\sim\left(\frac{t}{t_{0}}\right)^{-\frac{1}{2}\sum_{i=0}^{N-1}i}
\det B^{L}
\end{equation} 
where $(B^{L})_{nm}=H_{n-1}(x_m/2\sqrt{Dt_{0}})$, with $n,m=1,\ldots,N$.  
The determinant  $\det B^{L}$ is proportional to the Vandermonde 
determinant \cite{fisher}: $\det B^L = (Dt_0)^{-N(N-1)/2}\,
\prod_{i<j}|x_i-x_j|$, and  all  $t_{0}$-dependence 
drops out, as it must, to give the long-time behaviour 
\begin{equation}
Q({\bf x},t)\sim t^{-\frac{N(N-1)}{4}}.
\label{Fisher}
\end{equation} 
which  is just  the result  Fisher obtained  \cite{fisher}.  But our
approach  to the  problem also  gives  a simple  way to  obtain
expressions for the survival probability for $N$ vicious walkers with an
absorbing or reflecting wall at the origin (and all walkers starting on 
one side of the wall). 

The essential  arguments have been given in  the preceding subsection.
For  an absorbing  (reflecting) boundary,  only the  odd  (even) basis
functions   contribute.    Note   first   that   the   Fisher   result
(\ref{Fisher})  follows immediately  from (\ref{harmonic})  on setting
$a=1/2$ and $T = \ln(t/t_0)$.  The detailed discussion above was given
mainly to show how the arbitrary  scale $t_0$ drops out. To obtain the
asymptotic results for  a reflecting or absorbing wall  at the origin,
we can  simple make the same replacements  in Eqs.\ (\ref{reflecting})
and  (\ref{absorbing}) respectively.  For the  absorbing  boundary, we
recover the result of Krattenthaler et al.~\cite{krattenthaler}:
\begin{equation}
Q({\bf x},t)\sim t^{-\frac{N^{2}}{2}}\ \ ({\rm absorbing\ wall}),
\end{equation} 
while for a reflecting wall we obtain 
\begin{equation}
Q({\bf x},t)\sim t^{-\frac{N(N-1)}{2}}\ \ ({\rm reflecting\ wall}).
\end{equation} 
The latter is, to our knowledge, a new result. 

As  a final  comment we  note  that the  case where  the absorbing  or
reflecting wall moves,  with a displacement $x_w =  ct^{1/2}$, is also
amenable  in  principle to  exact  analysis.  The  change of  variable
(\ref{cov}) maps the problem to one where the $N$ walkers  move in  a
harmonic oscillator potential, and the absorbing or reflecting wall is
at a  {\em fixed  position} in the  new coordinates. This  problem has
been  analysed for  a  single walker  \cite{KrapivskyRedner}, and  the
survival  probability decays as $t^{-\theta}$, where the exponent 
$\theta$  is found  to vary  continuously  with the amplitude,  $c$,  
of  the  wall  displacement.  The  same  qualitative features will be 
present  for $N$ vicious walkers. For  a reflecting  (R) or
absorbing (A) boundary, one will obtain a decay exponent $\theta_{R,A}
=   N(N-1)/4  +   f_{R,A}(c,N)$,  where   $f_{R,A}(-\infty,N)   =  0$,
corresponding to a  rapidly receding wall, which will  be equivalent to
no  wall at  all, and  $f_R(0,N)  = N(N-1)/4$,  $f_A(0,N) =  N(N+1)/4$
correspond to a static wall.

\section{Conclusion}
In  this   paper  we  have derived  the exact  asymptotics   for  the  
survival probability of vicious walkers moving in a square well potential 
and a harmonic potential with various combinations of absorbing  and  
reflecting walls. The results for a harmonic potential have been used to 
find the properties of free vicious walkers (zero potential) through a 
change of variables, and a new result obtained for the case of a single 
reflecting boundary. Comparing  all results for each potential  one 
recognises that the survival probability decays faster when the number  
of walls is increased, with absorbing walls causing a faster decrease 
than reflecting walls, in accord with intuitive expectations.


\begin{thebibliography}{30}
\expandafter\ifx\csname natexlab\endcsname\relax\def\natexlab#1{#1}\fi
\expandafter\ifx\csname bibnamefont\endcsname\relax
  \def\bibnamefont#1{#1}\fi
\expandafter\ifx\csname bibfnamefont\endcsname\relax
  \def\bibfnamefont#1{#1}\fi
\expandafter\ifx\csname citenamefont\endcsname\relax
  \def\citenamefont#1{#1}\fi
\expandafter\ifx\csname url\endcsname\relax
  \def\url#1{\texttt{#1}}\fi
\expandafter\ifx\csname urlprefix\endcsname\relax\def\urlprefix{URL }\fi
\providecommand{\bibinfo}[2]{#2}
\providecommand{\eprint}[2][]{\url{#2}}

\bibitem[{\citenamefont{fisher}(1984)}]{fisher}
\bibinfo{author}{\bibfnamefont{M.~E.} \bibnamefont{Fisher}},
  \bibinfo{journal}{J. Stat. Phys.}\textbf{\bibinfo{volume}{34}},
  \bibinfo{pages}{667}
  (\bibinfo{year}{1984}).

\bibitem[{\citenamefont{huse and fisher}(1984)}]{husefisher}
\bibinfo{author}{\bibfnamefont{D.~A.} \bibnamefont{Huse}} \bibnamefont{and}
  \bibinfo{author}{\bibfnamefont{M. ~E.} \bibnamefont{Fisher}},
  \bibinfo{journal}{Phys. Rev. B}\textbf{\bibinfo{volume}{29}},
  \bibinfo{pages}{239}
  (\bibinfo{year}{1984}).

\bibitem[{\citenamefont{krattenthaler et al}(2000)}]{krattenthaler}
\bibinfo{author}{\bibfnamefont{C.} ~\bibnamefont{Krattenthaler}},
\bibinfo{author}{\bibfnamefont{A.~J.} \bibnamefont{Guttmann}},
\bibinfo{author}{\bibfnamefont{X.~G.} \bibnamefont{Viennot}},
  \bibinfo{journal}{J. Phys. A} \textbf{\bibinfo{volume}{33}},
  \bibinfo{pages}{8835}
  (\bibinfo{year}{2000}).

\bibitem[{\citenamefont{katori and tanemura}(2002)}]{katori}
\bibinfo{author}{\bibfnamefont{M.} ~\bibnamefont{Katori}} \bibnamefont{and}
\bibinfo{author}{\bibfnamefont{H.} ~\bibnamefont{Tanemura}},
  \bibinfo{journal}{Phys. Rev. E} \textbf{\bibinfo{volume}{66}},
  \bibinfo{pages}{011105}
  (\bibinfo{year}{2002}).

\bibitem[{\citenamefont{gardiner}(1985)}]{gardiner}
\bibinfo{author}{\bibfnamefont{C.~W.} \bibnamefont{Gardiner}},
  \emph{\bibinfo{title}{Handbook of stochastic methods for physics, chemistry and natural science}}
  (\bibinfo{publisher}{Springer}, \bibinfo{address}{Berlin, London},
  \bibinfo{year}{1985}, p. 128.

\bibitem[{\citenamefont{reichl}(1998)}]{reichl}
\bibinfo{author}{\bibfnamefont{L.~E.} \bibnamefont{Reichl}},
  \emph{\bibinfo{title}{A modern course in statistical physics}}
  (\bibinfo{publisher}{Wiley}, \bibinfo{address}{New York, Chichester},
  \bibinfo{year}{1998}), p. 274.

\bibitem[{\citenamefont{mapping}(1996)}]{map}
\bibinfo{author}{\bibfnamefont{S. ~N.} \bibnamefont{Majumdar}},
\bibinfo{author}{\bibfnamefont{C.} ~\bibnamefont{Sire}},
\bibinfo{author}{\bibnamefont{A.~J.} \bibnamefont{Bray}},
\bibnamefont{and}
\bibinfo{author}{\bibfnamefont{S.~J.} \bibnamefont{Cornell}},
  \bibinfo{journal}{Phys. Rev. Lett.} \textbf{\bibinfo{volume}{77}},
  \bibinfo{pages}{2867}
  (\bibinfo{year}{1996});
\bibinfo{author}{\bibfnamefont{B.} ~\bibnamefont{Derrida}},
\bibinfo{author}{\bibfnamefont{V.} ~\bibnamefont{Hakim}},
\bibnamefont{and}
\bibinfo{author}{\bibfnamefont{R.} ~\bibnamefont{Zeitak}},
\emph{ibid.},\bibinfo{pages}{2871}.

\bibitem[{\citenamefont{KrapivskyRedner}(1996)}]{KrapivskyRedner}
\bibinfo{author}{\bibfnamefont{P. ~L.} \bibnamefont{Krapivsky}},
\bibinfo{author}{\bibfnamefont{S.} ~\bibnamefont{Redner}},
 \bibinfo{journal}{Am. J. Phys.} \textbf{\bibinfo{volume}{64}},
  \bibinfo{pages}{546}
  (\bibinfo{year}{1996}).

\end{thebibliography}
\end{document}